# Optimized quantum entanglement network enabled by a state-multiplexing quantum light source

Yun-Ru Fan,[1,2,3] Yue Luo,[1,3] Kai Guo,[4,*] Jin-Peng Wu,[1,3] Hong Zeng,[1,3] Guang-Wei Deng,[1,3,5] You Wang,[1,6] Hai-Zhi Song,[1,6] Zhen Wang,[7] Li-Xing You,[7] Guang-Can Guo,[1,2,3,8] and Qiang Zhou[1,2,3,8,*]

[1]Institute of Fundamental and Frontier Sciences, University of Electronic Science and Technology of China, Chengdu 611731, China
[2]Center for Quantum Internet, Tianfu Jiangxi Laboratory, Chengdu 641419, China
[3]Key Laboratory of Quantum Physics and Photonic Quantum Information, Ministry of Education, University of Electronic Science and Technology of China, Chengdu 611731, China
[4]Institute of Systems Engineering, AMS, Beijing 100141, China
[5]Hefei National Laboratory, University of Science and Technology of China, Hefei 230088, China
[6]Southwest Institute of Technical Physics, Chengdu 610041, China
[7]National Key Laboratory of Materials for Integrated Circuits, Shanghai Institute of Microsystem and Information Technology, Chinese Academy of Sciences, Shanghai 200050, China
[8]CAS Key Laboratory of Quantum Information, University of Science and Technology of China, Hefei 230026, China

Corresponding author*. Email: guokai07203@hotmail.com; zhouqiang@uestc.edu.cn

## Abstract

A fully connected quantum network with a wavelength division multiplexing architecture plays an increasingly pivotal role in quantum information technology. With such architecture, an entanglement-based network has been demonstrated in which an entangled photon-pair source distributes quantum entanglement resources to many users. Despite these remarkable advances, the scalability of the architecture could be constrained by the finite spectrum resource, where $\mathcal{O}(N^2)$ wavelength channels are needed to connect $N$ users, thus impeding further progress in real-world scenarios. Here, we propose an optimized scheme for the wavelength division multiplexing entanglement-based network using a state-multiplexing quantum light source. With a dual-pump configuration, the feasibility of our approach is demonstrated by generating state-multiplexing photon pairs at multiple wavelength channels with a silicon nitride microring resonator chip. In our demonstration, we establish a fully connected graph between four users with six wavelength channels - saving half of which without sacrificing functionality and performance of the secure communication. A total asymptotic secure key rate of 1946.9 bps is obtained by performing the BBM92 protocol with the distributed state. The network topology of our method has great potential for developing a scalable quantum network with significantly minimized infrastructure requirements.

**Introduction**

Entanglement-based quantum network, in which the information is encoded on and measured from entangled photons, facilitates the development of quantum computation, quantum metrology, and quantum communication *(1-5)*. The quantum network has undergone extensive exploration, encompassing different types of configurations, such as point-to-point (*6*), trusted-node (*7-16*), point-to-multipoint (17-19), and fully connected network (*20-27*). Among these, the fully connected network, i.e., each user of the network simultaneously sharing quantum correlations and exchanging quantum information or quantum secure key with every other user, has emerged as one of the most versatile and robust architectures with the advancement of multiplexing technology in degrees of freedom in wavelength, space, and time. Harnessing progresses in wavelength division multiplexing (WDM), a four-user entanglement-based wavelength-multiplexed quantum network has been realized by using a polarization-entangled photon pair with twelve wavelength channels (*20*). In such a network, $N\times(N-1)$ wavelength channels are needed to accommodate *N* users, which is constrained to only a few ones due to the finite spectrum resource of the photon-pair source (*28-36*). To expand the network with available resources, a quadratic improvement in wavelength saving has been achieved with the establishment of an eight-user metropolitan network featuring 16 wavelength channels - with eight beamsplitters for passively temporal multiplexing at the expense of decreasing the rate of information (*21*). To further achieve a scalable and reconfigurable quantum network, flex-grid entanglement distributions have been implemented by employing multiplexing techniques based on wavelength selective switch technology, which enables the reconfigurability of a quantum network with ever-increasing complexity and service demands (*25-27*).

Despite these great promises, the fully connected quantum network has been demonstrated based on such a quantum light source that a quantum state is prepared on a wavelength. To leverage the finite wavelength resources, we propose a novel scheme of the optimized quantum network based on a state-multiplexing quantum light source. This approach employs two lasers to pump a third-order nonlinear optical device, facilitating three spontaneous four-wave mixing (SFWM) processes simultaneously - one non-degenerate (*37-43*) and two degenerate processes (*44-46*). Therefore, a wavelength overlap could be created, enabling photons at one wavelength channel to be correlated with photons at the other three wavelength channels, and thus reducing the wavelength resources needed for a fully connected network.

In this work, we developed a state-multiplexing photon pair source by using two lasers to pump a fiber-pigtailed silicon nitride microring resonator (MRR) chip with a free spectral range (FSR) of 200 GHz. The source enables a fully connected graph between four users with six wavelength channels, which saves half of the wavelength channels compared with the previous result (*20*). Our result shows that such a framework could allow simultaneous and secure connections with a total asymptotic secure key rate of 1946.9 bps by performing the BBM92 protocol. The network topology exhibits the potential to establish a scalable and reconfigurable quantum network, which reduces the infrastructure requirement, thus paving the way for developing a large-scale quantum secure network in future.

## Results

### Network scheme with state-multiplexing quantum light source

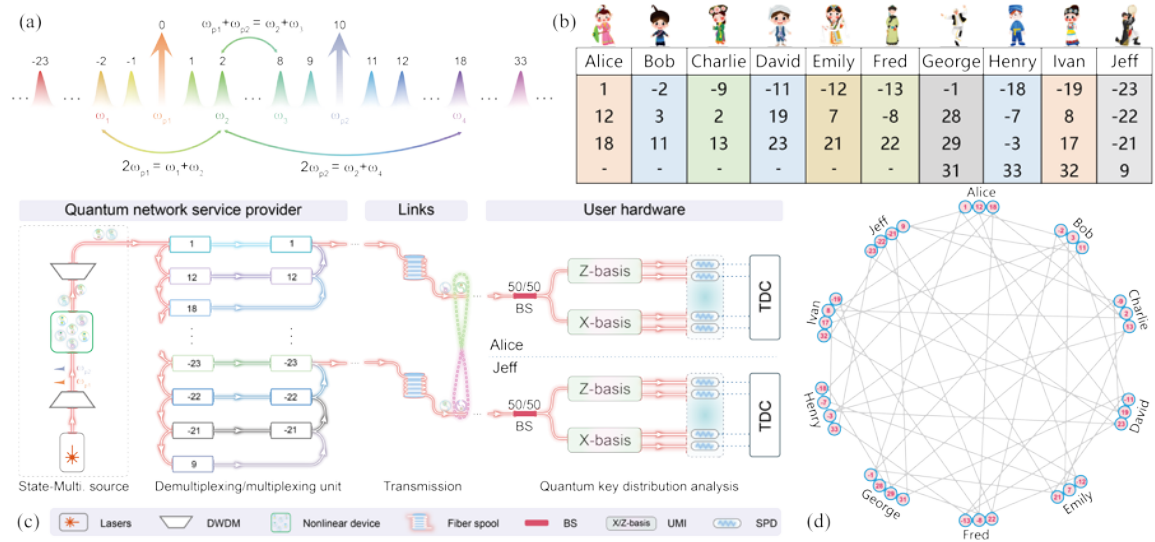

**Fig. 1 Schematic of optimized quantum network with a state-multiplexing quantum light source.** **(a)** State-multiplexing quantum light source with dual pump configuration, i.e., $2\omega_{p1}$, $2\omega_{p2}$, and $\omega_{p1}+\omega_{p2}$, respectively. Photons at the wavelength of $\omega_2$ can be correlated/entangled with photons at another three wavelengths - $\omega_1$, $\omega_3$, and $\omega_4$, i.e., three quantum states are prepared on a common wavelength by multiplexing. **(b)** Wavelength allocation of optimized quantum network for ten users. With pump photons labeled as 0 and 10, two users can establish connections by photon pair whose labels sum up to 0, 10, or 20, respectively. **(c)** Physical layer of the optimized quantum network. It consists of the quantum network service provider (QNSP), links, and user hardware, in which quantum states from QNSP are distributed to each user via fiber links. Each user is equipped with a beam splitter and two interferometers to perform the random choice of measurement basis between $Z$ (0 or $\pi$) and $X$ ($\pi/2$ or $3\pi/2$). **(d)** Communication layer of the optimized quantum network for ten users. With the optimized scheme, ten users are fully connected with 34 wavelength channels, which saves 56 wavelength channels compared to the previous scheme (*20*).

As shown in Fig. 1(a), state-multiplexing photon pairs are generated by pumping a nonlinear optical device with a laser module emitting light at two wavelengths. Taking the third-order nonlinear optical device as an example, the degenerate and non-degenerate SFWM processes could occur at the same time, i.e., correlated/entangled photons are generated with the pump configurations of $2\omega_{p1}$, $2\omega_{p2}$, and $\omega_{p1}+\omega_{p2}$, respectively. For instance, we denote the two wavelengths of the pump laser module as 0 and 10, whereby the photon pairs are entangled when the sum of their numerical labels equals 0, 20 or 10, corresponding to the degenerate or non-degenerate SFWM processes, respectively. For the wavelength at $\omega_2$ (labeled as 2), the photons could be correlated/entangled with those at another three wavelengths $\omega_1$, $\omega_3$, and $\omega_4$ - labeled as -2, 8, and 18. Thus, three quantum states are multiplexed on such a common wavelength with our method. Leveraging such a state-multiplexing quantum light source, a user who occupies one wavelength resource can connect with the other three users.

Physical and communication layers of the optimized quantum network featuring a state-multiplexing quantum light source are illustrated in Figs. 1(c) and 1(d), respectively. The physical layer contains a central quantum network service provider (QNSP), fiber links, and user hardware. As shown in Fig. 1(c), the QNSP includes the state-multiplexing quantum light source and the demultiplexing/multiplexing unit, while the user hardware consists of a beam splitter (BS), two unbalanced Michelson interferometers (UMIs), and four single-photon detectors (SPDs). Photons incident on the BS and UMIs, where they are measured in the Z basis (0 or $\pi$) or the X basis ($\pi/2$ or $3\pi/2$), thus enables the measurement in the diagonal/antidiagonal phase basis. The communication layer of the network facilitates a fully connected graph, enabling entanglement distribution, quantum information exchange, and secure communication between all pairs of users. We conceptually refer the ten users of our network as Alice (A), Bob (B), Charlie (C), Dave (D), Emily (E), Fred (F), George (G), Henry (H), Ivan (I), and Jeff (J). Every user within the ten-node network receives wavelength channels identified by numerical labels. The wavelength allocation is shown in Fig. 1(b). Benefiting from the state-multiplexing quantum light source, this scenario necessitates 34 wavelength channels for the ten-node network, a considerable reduction compared to 90 wavelength channels suggested in previous result (*20*).

**State-multiplexing quantum light source with dual pumps**

The generation and characterization of quantum state-multiplexing photon pairs with dual pumps is illustrated in Fig. 2. A fiber-pigtailed silicon nitride microring resonator chip is utilized with an FSR of ~200 GHz and a quality factor of ~$10^6$ (*47*). See more details in Methods and Supplemental Material Note1. As shown in Fig. 2(a), two independent continuous-wave tunable lasers (TLs) at the wavelengths of 1550.12 nm and 1540.56 nm without phase-stable with each other are used to generate photon pairs, which are multiplexed by using dense wavelength division multiplexers (DWDMs) with 200-GHz-spacing at the ITU channels of C34 and C46. In our experiments, the power of the two pump lasers is set to be equal in order to balance the efficiencies of the degenerate and non-degenerate type-0 SFWM processes, ensuring the optimized performance of the state-multiplexing quantum light source (*40, 48*). Quantum state-multiplexing photon pairs are generated in the resonator. After the pump rejection, signal and idler photons are selected by DWDMs and detected and recorded by superconducting nanowire single-photon detectors (SNSPDs) and time-to-digital converter (TDC), respectively, as shown in Fig. 2(b).

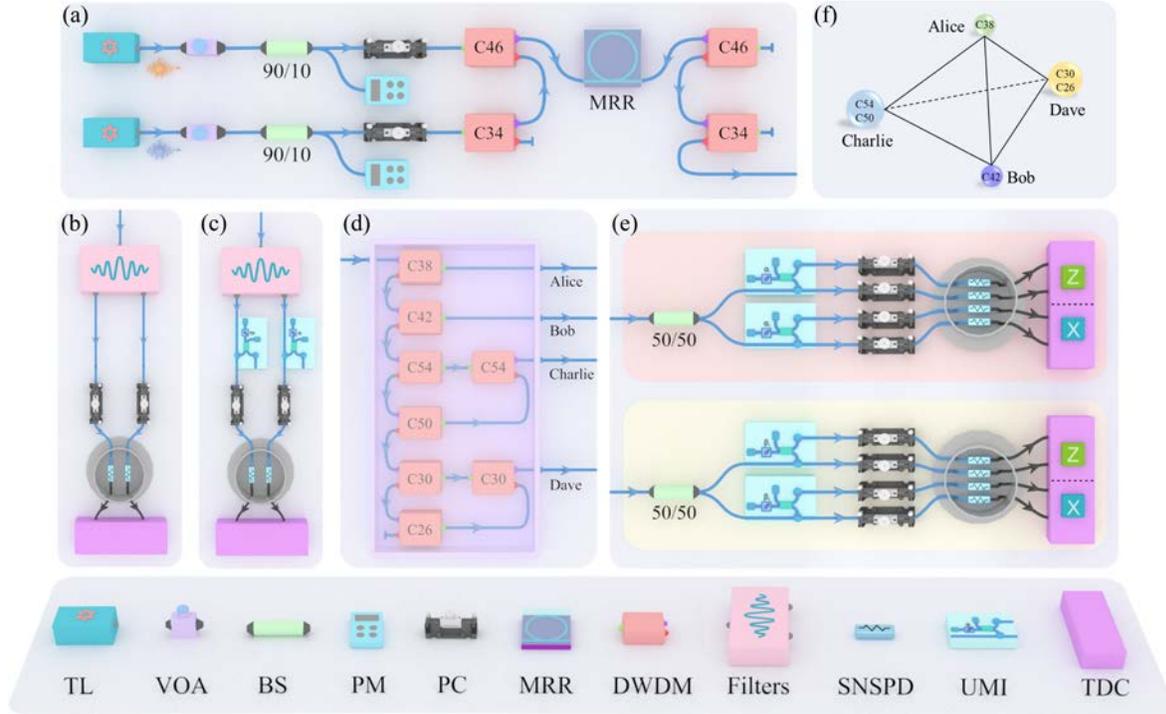

**Fig. 2 Schematic diagram of experimental setups.** (**a**) Generation of correlated photon pairs with dual pumps. (**b**) Correlation properties. (**c**) Energy-time entanglement with Franson interference. (**d**) Demultiplexing/multiplexing unit for the four-node quantum network with six wavelength channels. (**e**) Measurement setup for quantum key distribution using the BBM92 protocol. (**f**) Communication layer and wavelength allocation. Note that the sketch of the PC is adopted from Thorlabs. TL: Tunable laser; VOA: Variable optical attenuator; BS: Beam splitter; PM: Power meter; PC: Polarization controller; MRR: Microring resonator; DWDM: Dense wavelength division multiplexer; SNSPD: Superconducting nanowire single-photon detector; UMI: Unbalanced Michelson interferometer; TDC: Time-to-digital converter.

To characterize the quantum correlation property of generated photons, we select the ITU channel C38 as the common wavelength, photons from which are correlated/entangled with those from the ITU channels C30, C42, and C54. The single side count rates at C30, C42, and C54 with different pump power are measured as shown in Fig. 3(a). The error bars of the count rate are obtained by the Poissonian photon-counting statistic. The difference in the rates comes from the different quality factors and transmission losses of these channels. See more details in Supplemental Material Note2. The coincidence count rates and coincidence-to-accidental ratios (CARs) are given in Figs. 3(b) and 3(c). In Fig. 3(b), the inset shows the coincidence histograms for the correlated photon pairs between C30&C38, C42&C38, and C54&C38 at a pump power of 1 mW, respectively. It is worth noting that the coincidence count rates and CARs in the non-degenerate case are higher than those in the degenerate case because of the different efficiency between degenerate and non-degenerate SFWM processes (*37, 42, 43, 49*).

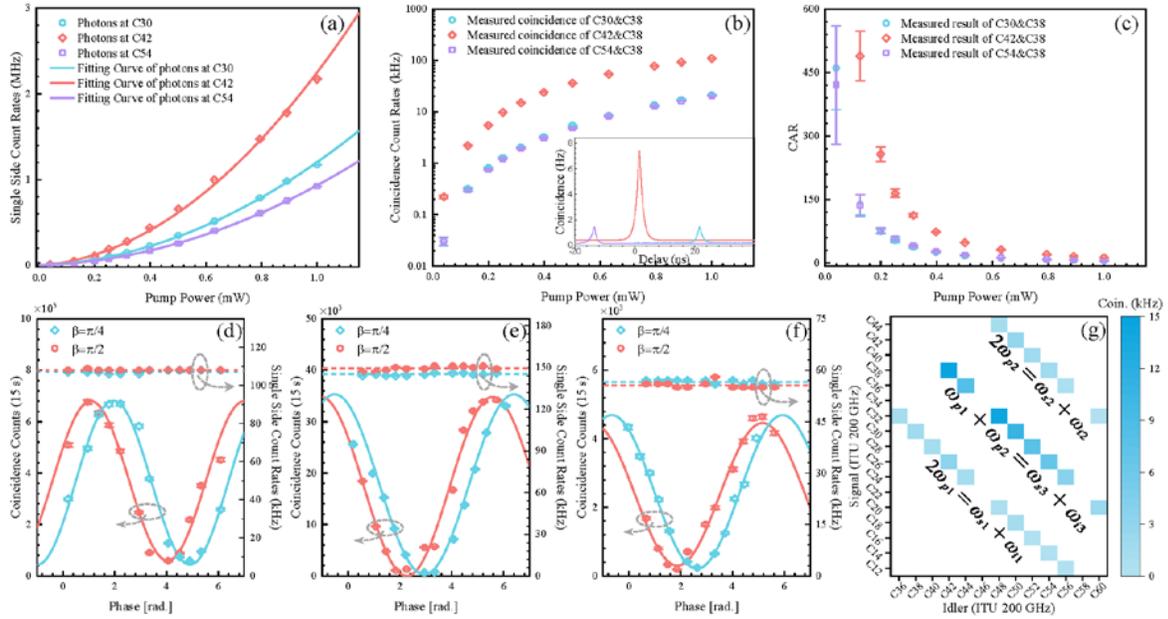

**Fig. 3 Experimental results of the state-multiplexing quantum light source with dual pumps.** (**a**) Single side count rates of photons at different levels of pump power at the wavelength of C30, C42, and C54, which is correlated/entangled with the photons at C38 - the state-multiplexing state. (**b**) Coincidence count rates and (**c**) CAR of photon pairs for C30&C38, C38&C42, and C38&C54. The inset of (b) gives the coincidence histograms for the three correlated photon pairs. (**d**)-(**f**) Results of Franson interference for C30&C38, C38&C42, and C38&C54, respectively. (**g**) Coincidence count rates of 24-wavelength-paired photon pairs for degenerate and non-degenerate SFWM processes at pump powers at 0.32 mW.

The property of energy-time entanglement of the state-multiplexing quantum light source is verified by the Franson interference (*50, 51*) by using two identical UMIs with an additional phase difference of $\alpha$ or $\beta$ as shown in Fig. 2(c). Note that the interferometers are stabled by the proportional-integral-derivative (PID) feedback control with reference light. See more details in Supplemental Material Note3. The measured interference curves of C30&C38, C38&C42, and C38&C54 are shown in Figs. 3(d)-3(f) the phases of $\beta = \pi/2$ and $\pi/4$. Circles are experimental results, while the lines are the fitting curves with a 1000-time Monte Carlo method. For $\beta = \pi/4$, the raw visibility is obtained as 87.1±0.5%, 98.2±0.2%, and 90.4±0.6% without subtracting the accidental coincidence counts. The single side count rates of photons at C30, C42, and C54 keep constant at ~108 kHz, ~141 kHz, and ~57 kHz, respectively, which indicates that there is no single-photon interference in the measurement. The quantum correlation and entanglement properties of other photon pairs are characterized by measuring the coincidence events between signal and idler photons at different wavelength channels, as illustrated in Fig. 3(g) and Table I. Note that the correlation and entanglement properties at C22 and C58 are not shown in the results due to the classical light generated from the stimulated four-wave mixing process (*52, 53*).

**Table 1.** Experimental results of the Franson interference for energy-time entanglement generated from the state-multiplexing quantum light source through degenerate and non-degenerate type-0 SFWM processes.

| SFWM Process | ITU channels | Visibility | ITU channels | Visibility |
|---|---|---|---|---|
| Degenerate | C32&C36 | 91.1±0.4% | C30&C38 | 87.1±0.5% |
|  | C28&C40 | 93.9±0.5% | C26&C42 | 83.9±0.5% |
|  | C24&C44 | 99.8±0.3% | C20&C48 | 83.4±0.9% |
|  | C18&C50 | 85.4±0.6% | C16&C52 | 87.0±1.6% |
|  | C14&C54 | 83.4±1.1% | C12&C56 | 85.4±1.9% |
|  | C44&C48 | 90.8±0.4% | C42&C50 | 91.9±0.4% |
|  | C40&C52 | 83.7±0.5% | C38&C54 | 90.4±0.4% |
|  | C36&C56 | 90.3±0.3% | C32&C60 | 88.1±0.8% |
| Non-degenerate | C36&C44 | 94.9±0.2% | C38&C42 | 98.2±0.2% |
|  | C32&C48 | 96.2±0.3% | C30&C50 | 94.3±0.2% |
|  | C28&C52 | 98.4±0.2% | C26&C54 | 95.7±0.1% |
|  | C24&C56 | 99.9±0.3% | C20&C60 | 91.4±0.5% |

**Quantum key distribution**

With the developed state-multiplexing quantum light source, we demonstrate the feasibility of the energy-time entanglement-based quantum key distribution network between four users with six wavelength channels. As shown in Fig. 2(f), Alice (A), Bob (B), Charlie (C), and Dave (D) are connected by photon pairs at C38, C42, C54/C50, and C30/C26, respectively. Figure 2(d) shows the details of the demultiplexing/multiplexing unit, in which six DWDMs are used for demultiplexing while two DWDMs are used for multiplexing, respectively. See more details in Tables S1, S3, and S4 in Supplemental Material for losses of the different schemes. We analyze the quantum bit error rate (QBER) and the asymptotic secure key rate (SKR) between two different users by performing the BBM92 protocol (*54*) as shown in Fig. 2(e). The results indicate that the performance between Alice and Bob is the best because of the high-efficiency photon generation and low-noise wavelength configuration, while the performance between Bob and Dave is the worst as photons at C30 are considered noises in this case.

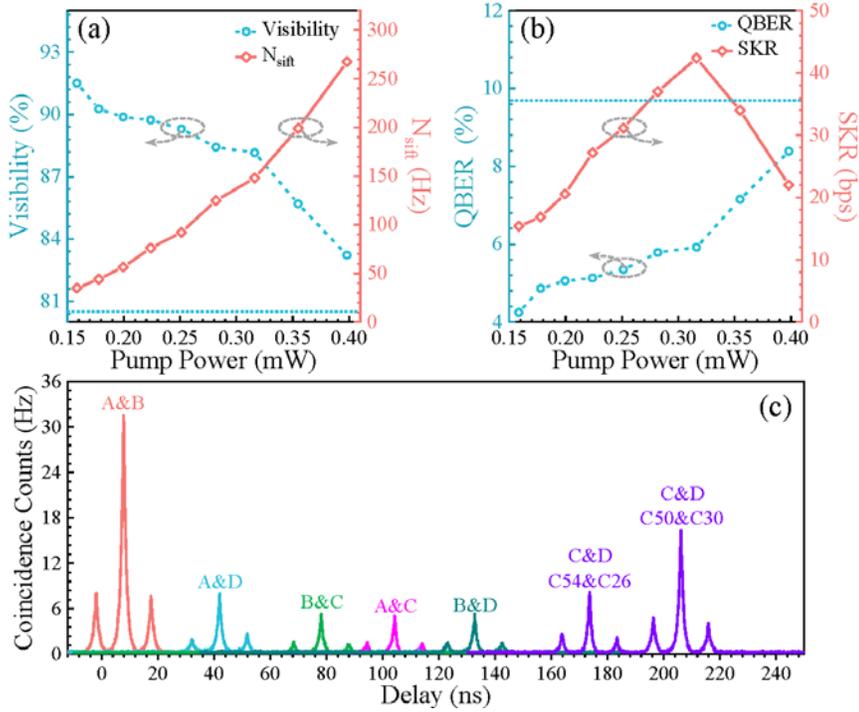

**Fig. 4 Experimental results of quantum key distribution with the developed state-multiplexing quantum light source. (a)** and **(b)** shows the visibility, $N_{sift}$, QBER, and SKR of B&D at different pump power levels. **(c)** Histograms between A&B, A&C, A&D, B&C, B&D, and C&D. Users C and D share two sets of correlated wavelengths with two sets of correlation peaks.

The secure key rate is calculated by $SKR = N_{sift} \times [1 - f(\delta_b) \times H_2(\delta_b) - H_2(\delta_p)]$ (22, 55-58), where $N_{sift}$ is the sifted key rate, $f(\delta_b)$ characterizes the efficiency of error correction with respect to Shannon's noisy coding theorem, the value of which is set to 1.2 in our calculation, $\delta_{b,p}$ is the bit or the phase error rate, and $H_2(\delta_{b,p})$ is the binary entropy function as $H_2(\delta_{b,p}) = -\delta_{b,p}\log_2(\delta_{b,p}) - (1-\delta_{b,p})\log_2(1-\delta_{b,p})$, respectively. In the experiments, $N_{sift}$ and QBER increase as the pump power, leading to a trade-off between $N_{sift}$ and QBER for a high-performance quantum network. To obtain the optimal pump power, we measure the visibility, $N_{sift}$, QBER, and SKR of B&D at different pump powers, as shown in Figs. 4(a) and 4(b) (55). At a pump power of 0.4 mW, the visibility and QBER approach the limit for successful key generation, which is 80.9% and 9.5% with $f(\delta_b) = 1.2$ respectively, as illustrated by the gray dashed lines in Figs. 4(a) and 4(b). See more details of the calculation in Supplemental Material Note7. Further increasing the pump power would lead to failure of key generation between users B and D. Therefore, we set the pump powers as 0.4 mW and analyze the properties between different combinations of users. The coincidence histogram results of A&B, A&C, A&D, B&C, B&D, and C&D are shown in Fig. 4(c). The results of quantum key distribution are listed in Table II. A total SKR of 1946.9 bps among four users is obtained, which is higher than that by using a single pump scheme. See more details in Table S5 in Supplemental Material for the performance comparison. The results suggest that our scheme not only preserves the functionality of secure communication but also conserves wavelength channels and enhances performance.

It is noted that C and D are connected by two entangled photon pairs - C50&C30 and C54&C26.

**Table 2. Results of the QBER and SKR between A&B, A&C, A&D, B&C, B&D, and C&D.**

| Users | ITU channel | $N_{sift}$ (Hz) | Visibility | QBER | SKR (bps) |
|---|---|---|---|---|---|
| A&B | C38&C42 | 1613.9 | 97.1% | 1.4% | 1230.6 |
| A&C | C38&C54 | 260.0 | 84.4% | 7.8% | 34.2 |
| A&D | C38&C30 | 405.6 | 87.1% | 6.5% | 97.6 |
| B&C | C42&C50 | 268.2 | 84.9% | 7.6% | 39.9 |
| B&D | C42&C26 | 267.4 | 83.2% | 8.4% | 22.9 |
| C&D | C50&C30 | 585.0 | 93.2% | 3.4% | 454.6 |
| C&D | C54&C26 | 460.0 | 84.8% | 7.6% | 67.1 |
| **Total** | | | | | 1946.9 |

**Discussion**

To further improve the SKR of the quantum network, we can utilize a quantum light source with a smaller coincidence window or coherence time (*59*), which can be realized by utilizing dual-Mach-Zehner microring device (*60*) and dual-microring device with parity-time symmetry (*61*). See more details for the theoretical analysis of SKR in Supplemental Material Note7. One may argue that the QBER and hence the SKR are limited by the inevitable noise with our scheme. For instance, in our demonstration, three states are prepared on one wavelength by multiplexing, with only one state being used while the other two are considered as noise. However, this is also the case in the previous scheme, where the quantum state in one wavelength channel is contaminated by other wavelength channels without demultiplexing or frequency-resolved detection at the user node. The QBER and SKR in the quantum network with state-multiplexing quantum light source are not only determined by the generation properties of the photon pairs, but also determined by the configuration of the network topology. For instance, in our demonstration, users C and D occupy two wavelength channels, i.e., C54/C50 and C30/C26, respectively. Therefore, the correlation properties of C38&C54 and C38&C30 are also influenced by photons in the wavelength channels of C50 and C26. One can balance the generation efficiencies of degenerate and non-degenerate processes and configure the arrangement of DWDMs among channels, and can also apply the flexible grid idea to dynamically balance the SKR (*25-27*).

The long-term goal of a full-fledged quantum internet requires a quantum communication network that supports the connection among as many users as possible using minimal wavelength channels. As the number of users increases, the advantages of our scheme become increasingly pronounced. The pump configuration can be customized and dynamically adjusted to tailor specific requirements. For instance, employing three or more pump lasers would further extend the spectrum range in which entangled two-photon states can be generated in the device. More multiplexed states would be prepared on a wavelength, thus increasing the scalability of the quantum network. To further configure the entanglement distribution network, one can adjust the powers and wavelengths of pump lasers to dynamically manipulate its connections. It is worth noting that with more pumps, more challenges and complexities in noise filtering and channel assignment will be introduced. Especially wavelength channels could be occupied by light generated from stimulated FWM processes. One could utilize the type-II SFWM process to develop the state-multiplexing quantum light source, in which the stimulated FWM can be further suppressed (*40, 48*). Besides, the bandwidth of quantum light source can be further extended by dispersion engineering of the SiN microring with inverse-design approach (*62*). Alternatively, nonlinear optical devices fabricated with different third-order nonlinear optical materials, such as GaN (*36*), AlGaAs (*63-65*), can also be utilized for developing broadband state-multiplexing quantum light source. Furthermore, the number of wavelength channels can be increased by reducing the FSR of the microring (*28, 31*).

In summary, we have successfully implemented an optimized energy-time entanglement-based quantum key distribution network using a state-multiplexing quantum light source for the first time, which confirms the enhanced feasibility and scalability of the proposed network architecture. Our state-multiplexing scheme can also be applied to quantum networks with various entanglement resources such as polarization entanglement and time-bin entanglement. Besides, combining with the time-sharing (*24*), BS multiplexing (*21*), and flex-gride (*25-27*), our scheme enables the realization of larger-scale quantum network.

## Materials and Methods

### Details of silicon nitride microring resonator chip

Silicon nitride ($Si_3N_4$) microring resonator chip can offer enhanced nonlinear effects by leveraging their resonant nature thanks to its ultralow optical loss and tailorable dispersion (*34*). In our experiments, we demonstrate an optimized quantum network using a microring resonator with a $Si_3N_4$ chip. The width-height cross-section of our microring resonators is designed as 1.8 $\mu m \times$ 0.8 $\mu m$ for the anomalous dispersion. The gap between the waveguide and the ring is 0.35 $\mu m$ with over-coupling for high-performance quantum light generation and emission. The microring chip is packaged by single-mode I/O fibers with an insertion loss of 3.2 dB, which is realized by using microlenses for precise mode matching and optical collimators for efficient coupling. The image of the fiber-pigtailed chip with thermal stability management is shown in Fig. S1(a). The transmission spectra of the microring from 1533 nm to 1558 nm are shown in Fig. S1(b). Figures S1(c) and (d) give the details at C46 and C34 with a quality factor of $1.29 \times 10^6$ and $1.22 \times 10^6$, respectively.

### Experimental setup

The experimental setup for the generation of photon pairs is shown in Fig. 2(a) in the main text. Two continuous-wave tunable lasers (Toptica CTL1550) at 1550.12 nm (ITU Channel C34) and 1540.56 nm (ITU Channel C46) are utilized. The power is adjusted by a variable optical attenuator (VOA) and is monitored by a 90:10 beam splitter (BS) and a power meter (PM). The polarization controller (PC) is used to control the polarization to align the TE00 mode of the waveguide. To suppress the sideband noise of the laser and the Raman photons generated in the fiber, two high-isolation (>120 dB) dense wavelength division multiplexers (DWDMs) at C46 and C34 are employed before the microring chip. Then the pump light at wavelengths of C34 and C46 are multiplexed and injected into the chip. At the output of the chip, the residual pump laser is rejected by two DWDMs with an isolation of >50 dB.

**Thermal stability**

To maintain the thermal stability of the microring resonator, a thermoelectric cooler and a thermistor are packaged at the bottom of the device, both of which are connected to a temperature controller with PID method. The temperature stability can be maintained within 0.001°C, which ensures the thermal stability of the microring resonator.

**Frequency stability**

To keep the pump lasers at the resonant wavelengths of the microring resonator, a wavelength meter (WS8-10, HighFinesse) with PID option is used to stabilize the wavelengths of the two narrow linewidth lasers (DLC CTL 1550, Toptica). The measured frequency shifts of the two lasers are shown in the inset of Figs. S8(a) and (b), which is within 1 MHz for both channels. Compared to the linewidth of ~200 MHz for the resonance, the pump laser can be stably tuned into the microring resonator. At the resonant wavelengths, the stability of our setups is further monitored by measuring the powers of the two pump lasers after passing through the microring resonator. As shown in Fig. S8(c), the results indicate that conditions are almost the same during the measurement.

**Acknowledgments.** This work was supported by Sichuan Science and Technology Program (Nos. 2022YFSY0061, 2022YFSY0062, 2022YFSY0063, 2023YFSY0062, 2023YFSY0058, 2023NSFSC0048), the National Natural Science Foundation of China (Nos. 62475039, 62405046, 92365106, 62105371), Innovation Program for Quantum Science and Technology (No. 2021ZD0300701).

**Conflict of interest.** The authors declare no competing interests.

**Data availability.** All data needed to evaluate the conclusions in the paper are present in the paper and/or the Supplementary Information. Additional data related to this paper may be requested from the authors.

**Contributions.** Qiang Zhou conceived and supervised the project. Yun-Ru Fan mainly carried out the experiment and collected the experimental data with help of other authors. Zhen Wang and Li-Xing You developed and maintained the SNSPDs used in the experiment. Yun-Ru Fan, Kai Guo, and Qiang Zhou analyzed the data and wrote the manuscript with inputs from all other authors. All authors have given approval for the final version of the manuscript.

[53] Dong, S. et al. True single-photon stimulated four-wave mixing. ACS Photonics 4, 746-753 (2017).

[54] Bennett, C. H., Brassard, G. & Mermin, N. D. Quantum cryptography without Bell's theorem. Physical Review Letters 68, 557-559 (1992).

[55] Gisin, N. et al. Quantum cryptography. Reviews of Modern Physics 74, 145 (2002).

[56] Ma, X. F., Fung, C. H. F. & Lo, H. K. Quantum key distribution with entangled photon sources. Physical Review A 76, 012307 (2007).

[57] Yin, J. et al. Satellite-to-ground entanglement-based quantum key distribution. Physical Review Letters 119, 200501 (2017).

[58] Yin, J. et al. Entanglement-based secure quantum cryptography over 1,120 kilometres. Nature 582, 501-505 (2020).

[59] Pelet, Y. et al. Operational entanglement-based quantum key distribution over 50 km of field-deployed optical fibers. Physical Review Applied 20, 044006 (2023).

[60] Wu, C. et al. Bright photon-pair source based on a silicon dual-Mach-Zehnder microring. Science China Physics, Mechanics & Astronomy 63, 220362 (2020).

[61] Chen, N. et al. Parity-time-symmetry-enabled broadband quantum frequency-comb generation. Physical Review A 110, 023714 (2024).

[62] Lucas, E. et al. Tailoring microcombs with inverse-designed, meta-dispersion microresonators. Nature Photonics 17, 943-950 (2023).

[63] Steiner, T. J. et al. Continuous entanglement distribution from an AlGaAs-on-insulator microcomb for quantum communications. Optica Quantum 1, 55-62 (2023).

[64] Steiner, T. J. et al. Ultrabright entangled-photon-pair generation from an AlGaAs-on-insulator microring resonator. PRX Quantum 2, 010337 (2021).

[65] Pang, Y. M. et al. A versatile chip-scale platform for high-rate entanglement generation using an AlGaAs microresonator array. Print at https://doi.org/10.48550/arXiv.2412.16360 (2024).

Page 15 of 15

# Supporting Information for

## Optimized quantum entanglement network enabled by a state-multiplexing quantum light source

Yun-Ru Fan, Yue Luo, Kai Guo, Jin-Peng Wu, Hong Zeng, Guang-Wei Deng, You Wang, Hai-Zhi Song, Zhen Wang, Li-Xing You, Guang-Can Guo, and Qiang Zhou

Kai Guo, Qiang Zhou.
E-mail: guokai07203@hotmail.com, zhouqiang@uestc.edu.cn

**This PDF file includes:**

Supporting text
Figs. S1 to S8
Tables S1 to S5
SI References

**Supporting Information Text**

**Note1. Details of silicon nitride microring resonator chip.** Figure S1(a) shows the details of fiber-pigtailed $Si_3N_4$ microring resonator chip with thermal stability management. Figures S1(b),(c), and (d) show the measured transmission spectra of the microring from 1533 nm to 1558 nm, the resonance at the ITU channel of C46 and C34 with a quality factor of $1.29 \times 10^6$ and $1.22 \times 10^6$, respectively.

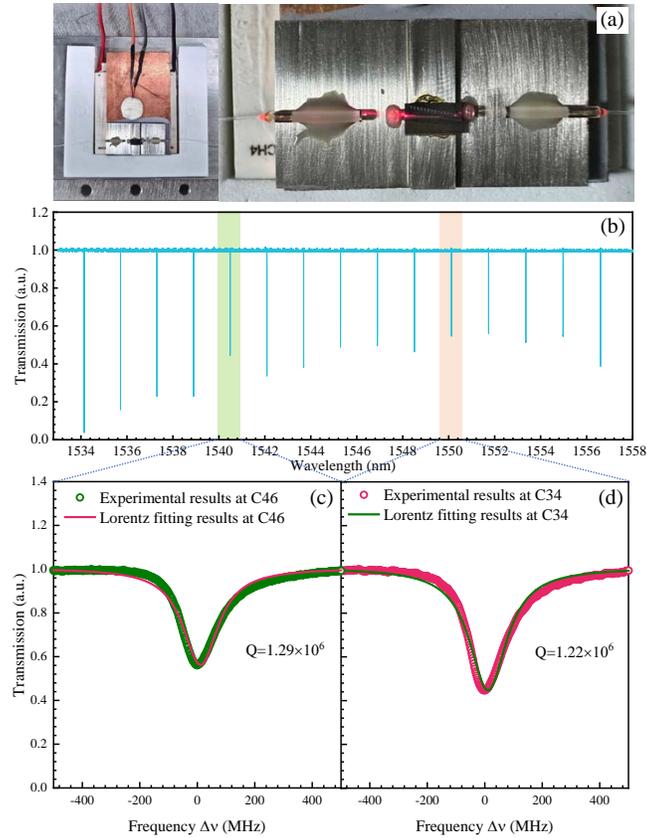

**Fig. S1.** Characterization of the $Si_3N_4$ microring chip. (a) Picture of the fiber-pigtailed $Si_3N_4$ device with thermal stability management. (b) Transmission spectrum from 1533 nm to 1558 nm. (c) Transmission spectrum C46 with a quality factor of $1.29 \times 10^6$. (d) Transmission spectrum C34 with a quality factor of $1.22 \times 10^6$.

**Yun-Ru Fan, Yue Luo, Kai Guo, Jin-Peng Wu, Hong Zeng, Guang-Wei Deng, You Wang, Hai-Zhi Song, Zhen Wang, Li-Xing You, Guang-Can Guo, and Qiang Zhou**

**Note2. Photon pairs generation with single- and dual-pump configuration.** We first measure the quantum correlation property with the single-pump configuration. As shown in Fig. 2(b) in the main text, the generated photon pairs, i.e., the signal and idler photons are selected by filters and detected by superconducting nanowire single-photon detectors (SNSPDs) with a detection efficiency of 85% and dark count rate of 50 Hz. The signals of SNSPD are sent to a time-to-digital converter (TDC) to record the coincidence events. With the single pump light at the wavelength of C34, the single side count rates on the resonance of C38 are shown in Fig. S2(a). The experimental results are illustrated by the circles, and the black line is the $aP_p^2 + bP_p$ fitting, in which the part of $aP_p^2$ is the contribution of correlated photons while $bP_p$ is the contribution of noise photons. Figure S2(b) shows the measured coincidence count rates, accidental coincidence count rates, and the ratios (CAR) between C38 and C30 with a coincidence window of 2.0 ns.

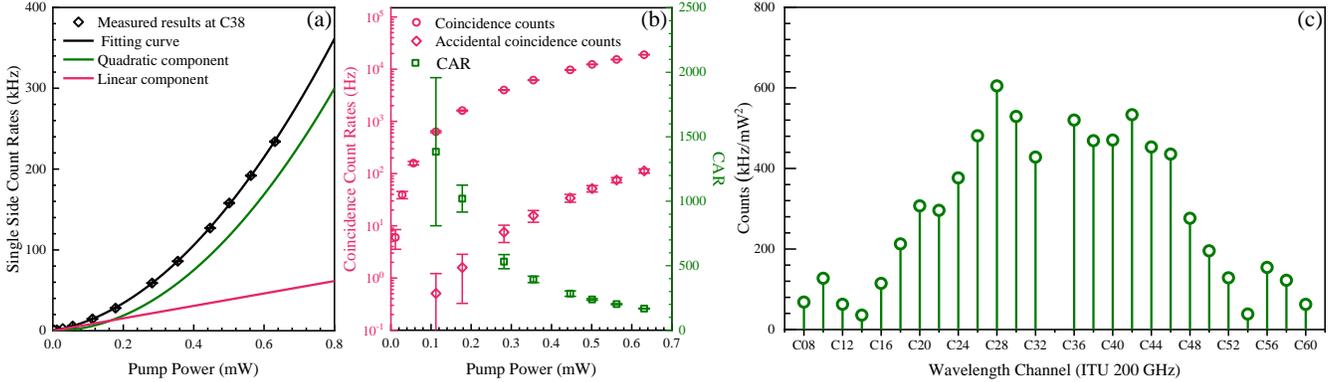

**Fig. S2.** Quantum correlation property of photon pairs with the single pump light at C34. (a) Single side count rates of C38 versus pump power. (b) Coincidence count rates, accidental coincidence count rates, and the coincidence-to-accidental ratio (CAR) between the photon pairs at C38 and C30 versus pump power. (c) Spectrum of the correlated photons.

To obtain the bandwidth of the spontaneous four-wave mixing process (SFWM), we further measure the single side count rates for different resonance wavelengths at different pump powers and calculate the contributions of the correlated photons by the $aP_p^2 + bP_p$ fitting as shown in Fig. S2(c). It can be seen that eight-wavelength-paired photon pairs are generated in a wavelength range of 25.6 nm, which is constrained by the phase-matching condition.

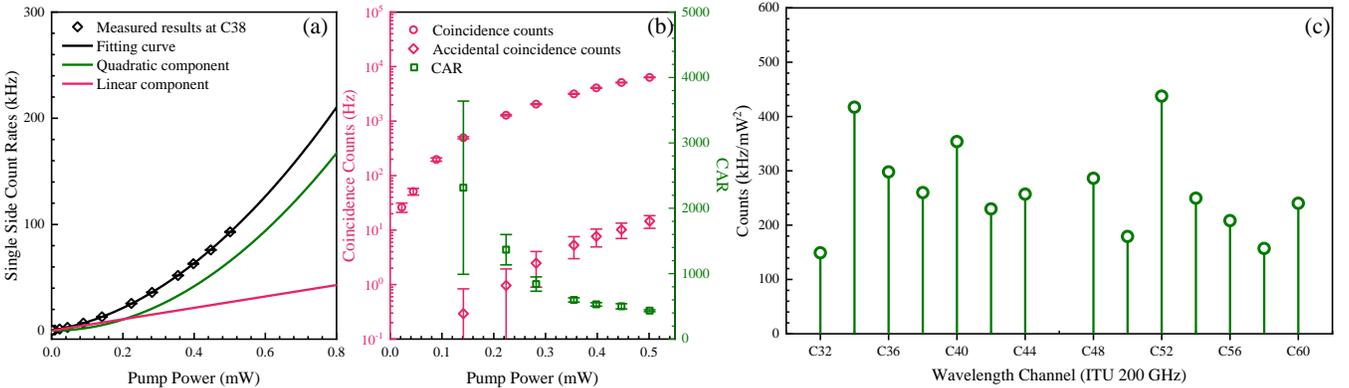

**Fig. S3.** Quantum correlation property of photon pairs with the single pump light at C46. (a) Single side count rates of C38 versus pump power. (b) Coincidence count rates, accidental coincidence count rates, and the CAR between the photon pairs at C38 and C54 versus pump power. (c) Spectrum of the correlated photons.

We also measure the quantum correlation property with the single pump light at the wavelength of C46 by using the same method. The single side count rates on the resonance of C38 are shown in Fig. S3(a), and the coincidence count rates, accidental coincidence count rates, and the CARs between C38 and C54 are given in Fig. S3(b). Figure S3(c) shows the contributions of correlated photon pairs at different wavelengths. Note that the other wavelengths are outside our measurements due to the lack of paired DWDMs.

To compare the properties between the non-degenerate and degenerate processes, we measure the quantum property with the dual-pump configuration, i.e., pump lights at C34 and C46 both input the $Si_3N_4$ chip. The single side count rates on the resonance of C38 are shown in Fig. S4(a) versus the pump powers. In our experiments, the pump power of C34 and C46 are the same. The coincidence count rates, accidental coincidence count rates, and the CARs are illustrated in Fig. S4(b). Utilizing the same approaches, the spectrum of the correlated photons is given in Fig. S4(c). It can be seen that the photon pair generation rate of the non-degenerate process is higher than that of the degenerate process, which is attributed to the

**Yun-Ru Fan, Yue Luo, Kai Guo, Jin-Peng Wu, Hong Zeng, Guang-Wei Deng, You Wang, Hai-Zhi Song, Zhen Wang, Li-Xing You, Guang-Can Guo, and Qiang Zhou**

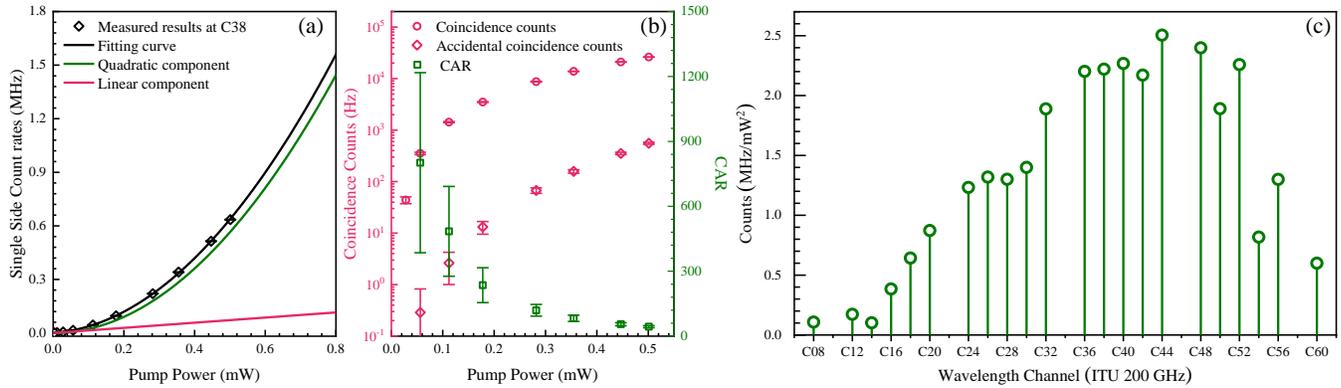

**Fig. S4.** Quantum correlation property of photon pairs with the dual pump lights at C34 and C46. (a) Single side count rates of C38 versus pump power. (b) Coincidence count rates, accidental coincidence count rates, and the CAR between the photon pairs at C38 and C42 versus pump power. (c) Spectrum of the correlated photons. Note that the contributions of correlated photon pairs include the non-degenerate and degenerate processes.

higher efficiency of the non-degenerate process. Besides, the non-degenerate spontaneous SFWM process is also limited by the dispersion and occurs in the same bandwidth of the degenerate process.

Furthermore, we measure the variety of CAR with the coincidence count rates by using the single- and dual-pump configuration, i.e., pump wavelength at C34, C46, and C34+C46, respectively. The results are shown in Fig. S5, which indicates lower CAR with dual-pump configuration due to the noise photons from the extra process. The Raman noise photons generated in the microring resonator need to be further reduced, and entangled two-photon states with higher generation rate could be achieved with a higher Q factor in the future.

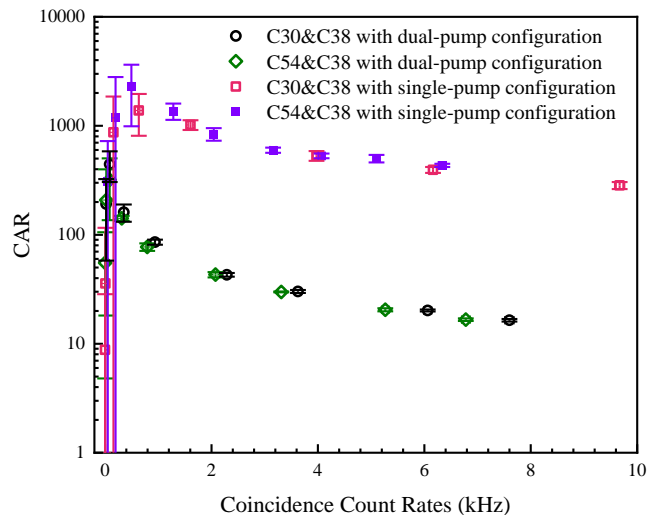

**Fig. S5.** Relationship between the CAR and coincidence count rates for single- and dual-pump configuration. With the dual-pump configuration, the property of photon pairs generated from the degenerate process degrades due to the photon pairs generated from the nondegenerate process are considered as "noise" and lead to more accidental coincidence events.

**Yun-Ru Fan, Yue Luo, Kai Guo, Jin-Peng Wu, Hong Zeng, Guang-Wei Deng, You Wang, Hai-Zhi Song, Zhen Wang, Li-Xing You, Guang-Can Guo, and Qiang Zhou**

**Note3. Quantum key distribution with single-pump configuration.** We first measure the performance of a four-user fully connected quantum key distribution network with single pump light at the wavelength of C46. Six pairs of the generated entangled photons at wavelengths ranging from C34 (1550.1 nm) to C58 (1531.1 nm) are selected and input the demultiplexing/multiplexing units, as shown in Fig. S6. The losses of the photons at different wavelengths in the units are listed in Table S1. These six pairs of photons are then distributed to four users: Alice (A), Bob (B), Charlie (C), and Dave (D). Each user receives three channels/wavelengths and shares an entanglement state with every other user in the network. It is worth noting that the quantum state is encoded in the degree of freedom of energy and time, which is immune to the dispersion-induced phase instability in the quantum key distribution(1). Besides, the dispersion caused broadening of coincidence window is negligible for the generated photon pairs with a spectral width of ∼200 MHz in our experiment.

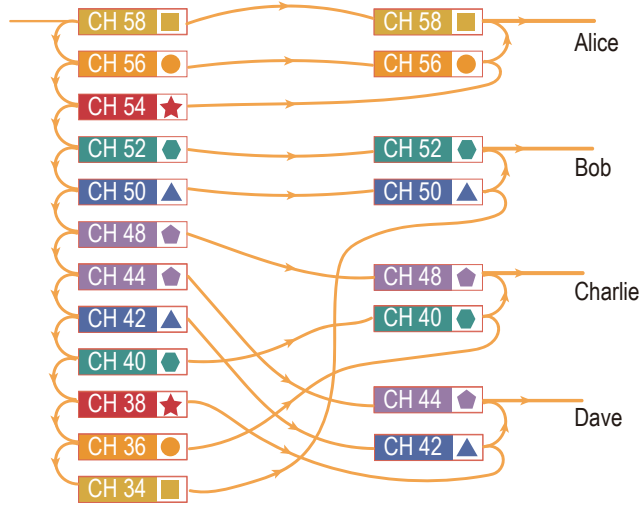

**Fig. S6.** Demultiplexing/multiplexing units. Six bipartite states are selected to create a fully connected network between four users.

**Table S1.** Losses of the demultiplexing/multiplexing unit at different wavelengths with single-pump configuration.

| Users | Channel | Wavelength (nm) | Loss (dB) |
|---|---|---|---|
| Alice | C58 | 1531.12 | 2.59 |
|  | C56 | 1532.68 | 3.35 |
|  | C54 | 1534.25 | 3.27 |
| Bob | C52 | 1535.82 | 3.26 |
|  | C50 | 1537.40 | 4.58 |
|  | C34 | 1550.12 | 5.66 |
| Charlie | C48 | 1538.98 | 4.00 |
|  | C40 | 1545.32 | 4.99 |
|  | C36 | 1548.51 | 5.30 |
| Dave | C44 | 1542.14 | 3.72 |
|  | C42 | 1543.73 | 5.33 |
|  | C38 | 1546.92 | 5.47 |

Yun-Ru Fan, Yue Luo, Kai Guo, Jin-Peng Wu, Hong Zeng, Guang-Wei Deng, You Wang, Hai-Zhi Song, Zhen Wang, Li-Xing You, Guang-Can Guo, and Qiang Zhou

As shown in Fig. S7, we use the BBM92 protocol to analyze the property of quantum key distribution between Alice and Bob. Alice/Bob splits their photons with a 50:50 beam splitter (BS), which performs the random choice of measurement basis between Z ($0/\pi$) and X ($\frac{\pi}{2}/\frac{3\pi}{2}$). It is worth noting that an attenuated continuous-wave laser is also injected into the unbalanced Michelson interferometer for phase stabilization based on feedback control. We characterize the performance of quantum key distribution between four users with six pairs of entangled photons - twelve wavelengths, as summarized in Table S2.

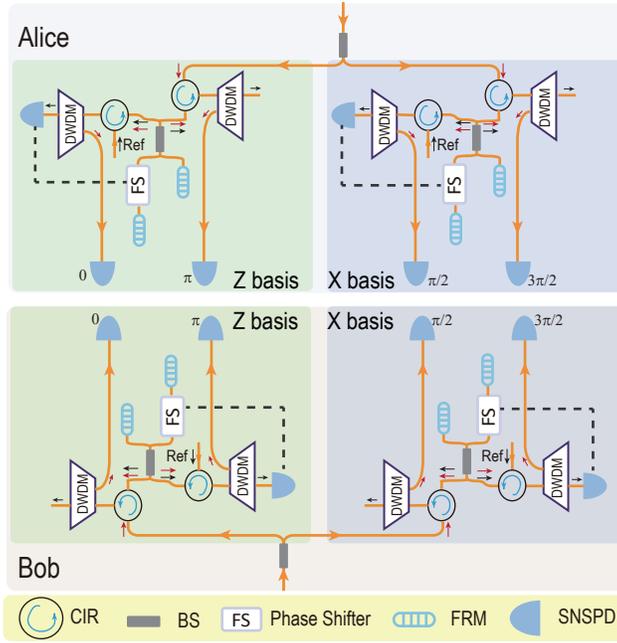

**Fig. S7.** Measurement setup for quantum key distribution using BBM92 protocol with feedback. The black arrows represent the direction of the reference light, and the red arrows represent the direction of the signal/idler photons.

**Table S2.** Measured results of quantum key distribution between four users.

| Users | $N_{sift}(Hz)$ | Visibility (%) | QBER (%) | SKR (bps) |
|---|---|---|---|---|
| A&B | 336.0 | 93.25 | 3.37 | 178.7 |
| A&C | 280.8 | 88.56 | 5.72 | 85.4 |
| A&D | 283.8 | 90.73 | 4.63 | 114.8 |
| B&C | 337.8 | 92.90 | 3.55 | 173.3 |
| B&D | 298.7 | 86.54 | 6.73 | 65.0 |
| C&D | 333.5 | 92.51 | 3.74 | 164.4 |
| Total | | | | 781.6 |

**Yun-Ru Fan, Yue Luo, Kai Guo, Jin-Peng Wu, Hong Zeng, Guang-Wei Deng, You Wang, Hai-Zhi Song, Zhen Wang, Li-Xing You, Guang-Can Guo, and Qiang Zhou**

**Note4. Losses with dual-pump scheme.** With dual-pump configuration, we use eight DWDMs in the demultiplexing/multiplexing unit as shown in Fig. 2(d). The losses of each channel and each component are shown in Table S3 and Table S4, respectively.

**Table S3.** Losses of the demultiplexing/multiplexing unit at different wavelengths with dual-pump configuration.

| Users | Channel | Wavelength (nm) | Loss (dB) |
|---|---|---|---|
| Alice | C38 | 1546.92 | 2.04 |
| Bob | C42 | 1543.73 | 2.35 |
| Charlie | C54 | 1534.25 | 3.19 |
|  | C50 | 1537.40 | 3.72 |
| Dave | C26 | 1556.55 | 2.39 |
|  | C30 | 1553.33 | 3.24 |

**Table S4.** Losses of different components

| Components | Loss (dB) | Efficiencies |
|---|---|---|
| Coupling | 1.5 | 71% |
| DWDM-C46 | 0.8 | 83% |
| DWDM-C34 | 0.7 | 85% |
| PC1 | 0.1 | 98% |
| PC2 | 0.1 | 98% |
| PC3 | 0.2 | 96% |
| PC4 | 0.1 | 98% |
| SNSPD1 | 1.0 | 79% |
| SNSPD2 | 1.0 | 80% |
| SNSPD3 | 1.3 | 74% |
| SNSPD4 | 0.9 | 82% |

**Yun-Ru Fan, Yue Luo, Kai Guo, Jin-Peng Wu, Hong Zeng, Guang-Wei Deng, You Wang, Hai-Zhi Song, Zhen Wang, Li-Xing You, Guang-Can Guo, and Qiang Zhou**

**Note5. Performance comparison of quantum key distribution based on the BBM92 protocol.** The performance comparison of a fully connected quantum key distribution network based on BBM92 protocol with multiple users is listed in Table S5.

**Table S5.** Performance comparison of a fully connected quantum key distribution network based on BBM92 protocol.

| Reference | Basis | Users | Wavelengths | SKR(bps) |
|---|---|---|---|---|
| Ref(2) | Polarization | 4 | 12 | 3~15 |
| Ref(3) | Polarization | 8 | 16 | 58~304 |
| Ref(4) | Energy-time | 4 | 12 | 58~252 (1298 in total) |
| Our work | Energy-time | 4 | 6 | 22~1230 (1946 in total) |

**Yun-Ru Fan, Yue Luo, Kai Guo, Jin-Peng Wu, Hong Zeng, Guang-Wei Deng, You Wang, Hai-Zhi Song, Zhen Wang, Li-Xing You, Guang-Can Guo, and Qiang Zhou**

**Note6. Stability of the experimental system.** Figures S8(a) and (b) show the frequency stability of the two lasers, indicating that the frequency shifts are within 1 MHz. After passing through the microring resonator, the output power at the resonant wavelengths is shown in Fig. S8(c), which demonstrates the stability of our setups.

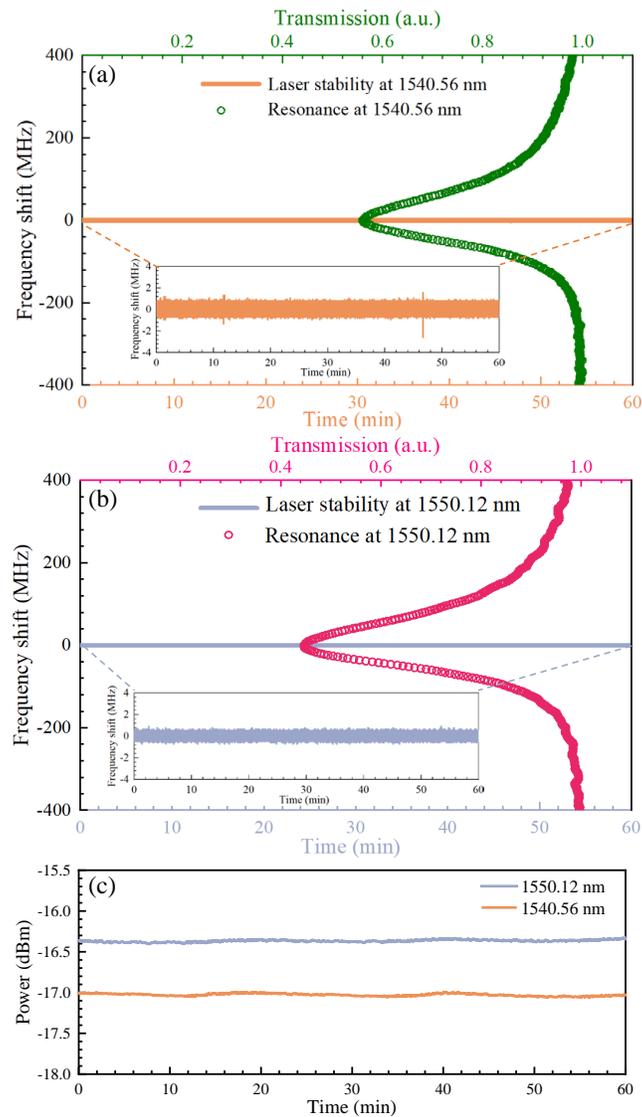

**Fig. S8.** Frequency shifts of pump lasers compared with transmissions of the microring resonator at (a) 1540.56 nm, and (b) 1550.12 nm, respectively. (c) Measured powers of the two pump lasers at the resonant wavelengths after passing through the microring resonator.

Yun-Ru Fan, Yue Luo, Kai Guo, Jin-Peng Wu, Hong Zeng, Guang-Wei Deng, You Wang, Hai-Zhi Song, Zhen Wang, Li-Xing You, Guang-Can Guo, and Qiang Zhou

**Note7. Theoretical analysis of SKR.** The performance of a quantum light source could be characterized by coherence time or the width of the coincidence window $\Delta\tau$ and the mean photon number $\bar{n}$ per coincidence window. The coincidence rates $C$ and the accidental coincidence count rates $A$ can be written as

$$C = \bar{n}\frac{1}{\Delta\tau} \qquad [1]$$

$$A = \bar{n}^2\frac{1}{\Delta\tau} \qquad [2]$$

Coincidence-to-accidental ratio (CAR), which is used to measure the signal-to-noise ratio of the quantum light source, can be calculated by the ratio between the coincidence count rates and the accidental count rates,

$$CAR = \frac{C}{A} = \frac{1}{\bar{n}} \qquad [3]$$

The visibility $V$ of Franson interference without subtraction of accidental coincidences is limited by the number of coincidences in the interference maximum, i.e., $C$ and the number of accidental ones, i.e., $A$. The visibility can be expressed as

$$V = \frac{C - A}{C + A}. \qquad [4]$$

In quantum entanglement distribution networks with BBM92 protocol, the SKR can be expressed as(4–8),

$$SKR \geqslant N_{sift} \times [1 - f(\delta_b) \times H_2(\delta_b) - H_2(\delta_p)] \qquad [5]$$

where $N_{sift}$ is the sifted key rate with $N_{sift} = C + A$, $f(\delta_b)$ characterizes the error correction efficiency with respect to Shannon's noisy coding theorem, $\delta_{b,p}$ is the bit or the phase error rate respectively for the X- and Z-basis measurements, and $H_2(\delta_{b,p})$ is the binary entropy function. For the energy-time entanglement, the X- and Z-basis measurements are symmetric, resulting in $\delta_b = \delta_p = x$, where $x$ is the overall QBER. In the experiments, $x$ can be calculated by

$$x = QBER = \frac{A}{C + A} = \frac{1 - V}{2} = \frac{1}{CAR + 1}, \qquad [6]$$

Therefore, $H_2(\delta_{b,p})$ can be expressed as $H_2(x) = -x\log_2 x - (1-x)\log_2(1-x)$. In our calculation, the value of $f(\delta_b)$ is set to 1.2 following the approach in the reference of (8). Then, the SKR can be expressed as

$$\begin{aligned} SKR &\geqslant N_{\text{sift}} \times [1 - 2.2H_2(x)] \\ &\geqslant (C + A) \times \{1 - 2.2[-x\log_2(x) - (1-x)\log_2(1-x)]\} \\ &\geqslant \frac{1-x}{\Delta\tau \times x^2} \times \{1 - 2.2[-x\log_2(x) - (1-x)\log_2(1-x)]\} \end{aligned} \qquad [7]$$

Therefore, the SKR depends on the property of the quantum light source, i.e., the coincidence count rates and the CAR with a given coincidence window. To further improve the SKR of the entanglement-based QKD network, a quantum light source with a smaller coincidence window or coherence time can be utilized, which can be realized by utilizing dual-Mach-Zehner microring device(9) and dual-microring device with parity-time symmetry(10). Besides, with a certain quantum light source, the SKR can be optimized with CAR(11).

**Yun-Ru Fan, Yue Luo, Kai Guo, Jin-Peng Wu, Hong Zeng, Guang-Wei Deng, You Wang, Hai-Zhi Song, Zhen Wang, Li-Xing You, Guang-Can Guo, and Qiang Zhou**